\documentclass{aa}  

\usepackage{graphicx}
\usepackage{txfonts}
\usepackage{ulem}
\usepackage{xcolor}
\definecolor{mycitelinkcolor}{rgb}{0.24705882, 0.27843137, 0.53333333}
\definecolor{myurlcolor}{rgb}{0.267004, 0.004874, 0.329415}
\usepackage[colorlinks=true, linkcolor=mycitelinkcolor, urlcolor=myurlcolor, citecolor=mycitelinkcolor]{hyperref}

\begin{document} 

\title{On the cosmology dependence of\\the cluster weak-lensing mass bias}
    
\author{S.~Bocquet \inst{1}\thanks{sebastian.bocquet@physik.lmu.de}
        \and
        A.~Fumagalli\inst{1, 2}
        \and
        C.~T.~Davies \inst{1}
        \and
        K.~Dolag\inst{1, 3}
        \and
        S.~Grandis \inst{4}
        \and
        J.~J.~Mohr \inst{1}
        }

\institute{
    University Observatory, Faculty of Physics, LMU Munich, Scheinerstr. 1, 81679 Munich, Germany 
    \and
    INAF-Osservatorio Astronomico di Trieste, Via G. B. Tiepolo 11, 34143 Trieste, Italy
    \and
    Max-Planck-Institut für Astrophysik, Karl-Scharzschild-Str. 1, 85748 Garching, Germany
    \and
    Universit\"at Innsbruck, Institut f\"ur Astro- und Teilchenphysik, Technikerstr. 25/8, 6020 Innsbruck, Austria
    }

\titlerunning{Cosmology dependence of the cluster weak-lensing mass bias}
\authorrunning{Bocquet, Fumagalli, Davies, Dolag, Grandis, and Mohr}

\date{Accepted 9 June 2026}

\abstract
  {Measurements of the shear induced by weak gravitational lensing around galaxy cluster lines of sight are the gold standard for calibrating cluster observable--mass relations, thereby enabling a robust and precise inference of cosmological parameters. The ``weak-lensing mass bias'' is the systematic offset between the true halo mass and the mass that is inferred from the lensing data using an imperfect model for the halo mass distribution.}
  {We study the impact of cosmology on the lensing mass bias to inform future cosmological analyses of galaxy clusters.}
  {We create synthetic lensing shear maps for 115\,920~projections of clusters with $M_{200\mathrm c}>1.56\times10^{14}\,h^{-1}M_\odot$ in a suite of \textsc{Magneticum} simulations. The simulation boxes are $896\,h^{-1}$Mpc on a side and are set up with 15 different combinations of the cosmological parameters $\Omega_\mathrm{m}$, $\Omega_\mathrm{b}$, $\sigma_8$, and $H_0$. Assuming a Navarro--Frenk--White profile, we extract ``weak-lensing mass'' measurements and quantify their bias $b_\mathrm{WL}$ with respect to the true halo mass. To investigate the impact of baryonic effects, we perform the analysis on gravity-only simulations and on their full-physics hydrodynamical counterparts.}
  {We confirm that assuming a fixed halo concentration or a fixed concentration--mass relation leads to cosmology-dependent changes of the mass bias. We report changes of up to $\Delta\ln b_\mathrm{WL}=0.030$ with respect to the bias obtained at the fiducial WMAP7 cosmology. Adopting a model for the concentration that also depends on cosmology absorbs the changes in halo profiles and we recover essentially constant values for the mass bias. Our analysis of hydrodynamical simulations suggests that future, more accurate models will also need to explicitly account for the strength of baryonic effects.}
  {The variation of the cluster weak-lensing mass bias over the range of cosmological parameters we probe here is small compared to the overall systematic error budget in current cluster lensing analyses. Nonetheless, we recommend the use of a model for halo concentration that explicitly depends on cosmology.}


\maketitle

\section{Introduction}
Galaxy clusters, by residing at the nodes of the cosmic web, trace the most massive collapsed peaks in the matter distribution. As the endpoints of hierarchical structure formation, they provide a unique window into the properties of the large-scale structure of the Universe \citep[see, e.g.,][]{Haiman01, Allen:2011zs,Kravtsov:2012zs}. Their number density and spatial distribution trace the growth of structure and the matter content of the Universe, making galaxy clusters a powerful cosmological probe that delivers competitive constraints on the matter density, the amplitude of matter fluctuations, and the nature of dark energy, thereby also enabling tests of the standard model and extensions thereof \citep{mantz15WtG, Planck:2015lwi, SPT:2018njh,DES:2018crd,DES:2021wwk,Salvati:2021gkt,Chiu:2022qgb,Fumagalli:2023yym,SPT:2024qbr,Ghirardini:2024yni,mazoun25, vogt25, Lesci:2025bqt, salcedo25, Sarieddine26}. This capability is being significantly enhanced in the era of surveys such as those conducted by eROSITA\footnote{\url{https://www.mpe.mpg.de/eROSITA}} \citep{Predehl2021A&A...647A...1P}, \textit{Euclid}\,\footnote{\url{https://sci.esa.int/euclid/}} \citep{Euclid:2025Mellier}, the Vera C. Rubin Observatory\footnote{\url{https://www.RubinObservatory.org/}} \citep{Ivezic219}, SPT-3G\footnote{\url{https://pole.uchicago.edu/}} \citep{sobrin22}, and the Simons Observatory\footnote{\url{https://simonsobservatory.org/}} \citep{SO19sciencegoals} that will deliver cluster catalogs of unprecedented size, substantially boosting the statistical constraining power of cosmological analyses. Fully exploiting these rich datasets, however, demands rigorous control over systematic uncertainties. 

In this context, the main limitation in current cluster cosmology lies in calibrating cluster masses, which underpins the accuracy of any cosmological inference. Direct mass measurements of clusters using gravitational lensing within a survey context are typically imprecise, and therefore individual cluster masses are typically inferred through secondary quantities that serve as mass proxies. Reliable mass calibration requires robust links between these observable cluster properties -- such as optical richness, X-ray luminosity, or the thermal Sunyaev--Zeldovich effect -- and the underlying halo mass, making it a critical step in extracting cosmological information from cluster surveys \citep[e.g.,][]{Kravtsov:2012zs, Pratt:2019cnf}.
Weak gravitational lensing has emerged as the most promising tool for this task. By measuring the coherent distortions imprinted on background galaxies by foreground clusters, weak lensing offers a way to estimate cluster masses that does not rely on assumptions about the dynamical or thermal state of the cluster \citep[e.g.,][]{Bartelmann:1999yn, Hoekstra:2013via}. As a result, lensing-based measurements are widely used to calibrate the mass--observable relations employed in the X-ray \citep{Applegate:2012kr, vonderlinden14WtG, Simet17ReflexBCS, grandis24, Chiu:2025cjh}, the thermal Sunyaev--Zeldovich effect \citep{vonderlinden14Planck, schrabback18, dietrich19, Nagarajan:2018ajl, Miyatake:2018lpb, zohren22, bocquet24a, shin25}, and the optically \citep{SDSS:2007xgc, DES:2016opl, Simet17SDSS, Bellagamba:2018gec,DES:2018kma, Lesci:2025bqt} selected cluster samples.

Nonetheless, weak-lensing mass estimates are not immune to systematic effects. Imperfect shape measurements, intrinsic alignments in populations of background galaxies, shear calibration uncertainties, photometric redshift errors, contamination by cluster members, miscentering of the cluster position, halo triaxiality and its coupling with halo orientation along the line of sight, and projections of large-scale structure all leave imprints on the measured shear signal. Besides amplifying the scatter of weak-lensing mass estimates around the true halo mass, these effects can lead to systematic offsets in the inferred lensing mass, introducing what is commonly referred to as the ``cluster weak-lensing mass bias''.

Weak-lensing masses are generally found to be modestly biased. Studies based on numerical simulations indicate that weak-lensing masses inferred by fitting spherically symmetric Navarro--Frenk--White \citep[NFW,][]{Navarro:1995iw} profiles to reduced tangential shear are typically underestimated by about 5--10\% \citep{Becker:2010xj}. This level of bias has been consistently confirmed by subsequent numerical analyses, which attribute it to departures from spherical symmetry, triaxiality, and line-of-sight structure \citep[e.g.,][]{Oguri:2011vj,Bahe:2011cb,Euclid:2023Giocoli}. However, in non-standard selection regimes the bias can increase significantly: for instance, for clusters identified as peaks in weak-lensing mass maps -- namely shear-selected systems -- the inferred masses are found to be biased high by $\sim 55\%$ on average, due to noise and projection effects that preferentially up-scatter lower-mass halos \citep{Chen:2019cjt}.

The shear that massive halos cause via gravitational lensing depends on cosmology. We distinguish between two effects:
\begin{itemize}
  \item Geometry: The lensing efficiency, given by the inverse critical surface mass density
  \begin{equation} \label{eq:Sigma_crit}
    \Sigma_{\rm crit}^{-1} = \frac{4\pi G}{c^2}\,\frac{D_{\rm l}}{D_{\rm s}}\,\mathrm{max}\left(0,  D_{\rm ls}\right),
  \end{equation}
  depends on the angular diameter distances
  $D_{\rm l}$ and $D_{\rm s}$ from the observer to the lens and the observer to the source respectively, and the angular diameter distance from the lens to the source $D_{\rm ls}$ (if the source is in front of the lens, then $D_{\rm ls}<0$ and there is no lensing effect).
  The critical surface mass density thus depends on the cosmological parameters that determine the geometry of the Universe (the densities $\Omega$ and the dark energy equation of state parameter $w(z)$). Accounting for this cosmological dependency is trivial, and cluster cosmology analyses that rely on cluster lensing have explicitly modeled this effect.
  \item Halo properties: Cosmological parameters that alter the history of structure formation are expected to leave an imprint on the halos themselves by changing their formation and growth history.
\end{itemize}
The goal of this paper is to investigate the impact of the latter effect -- the change of the cluster weak-lensing mass bias due to cosmology-dependent changes in the halo profiles --  while accounting for the cosmology-dependence of the lens--source geometry.
In Sect.~\ref{sec:theory}, we review the theoretical framework for weak gravitational lensing by massive halos. Sect.~\ref{sec:simulations} describes the simulations used in our work, while Sect.~\ref{sec:methods} outlines our analysis methodology. In Sect.~\ref{sec:results}, we present our results, and in Sect.~\ref{sec:conclusions}, we summarize our findings and discuss their implications.

\section{Theory} \label{sec:theory}
Weak gravitational lensing probes the projected mass distribution of foreground halos through the coherent distortion that their gravitational potential induces on the shapes of background galaxies. The observable quantity is the reduced tangential shear
\begin{equation} \label{eq:reduced_shear}
  g_\mathrm{t}(\mathbf x) = \frac{\gamma_\mathrm{t}}{1 - \kappa} (\mathbf x),
\end{equation}
with the convergence $\kappa$ and the tangential shear $\gamma_\mathrm{t}$.

To model $g_\mathrm{t}$, for convenience, we assume a spherically symmetric density distribution $\rho(r)$ that follows the NFW profile.\footnote{Alternatively, we could consider other profiles or extensions of the NFW profile \citep[e.g.,][]{Einasto1965, BaltzMarshallOguri2009}. While the numerical values of our results would likely differ, we do not expect our conclusions to change in a qualitative way.} Assuming a model for the halo concentration $c_{200\mathrm c}$ yields the density profile as a two-parameter family in halo mass $M_{200\mathrm c}$ and redshift. With the projected mass density $\Sigma_\mathrm{NFW}(R)$ and the excess surface mass density $\Delta\Sigma_\mathrm{NFW}(R)\equiv\langle\Sigma_\mathrm{NFW}(<R)\rangle-\Sigma_\mathrm{NFW}(R)$ at projected radius $R$, we model the shear profile as
\begin{equation} \label{eq:NFW_shear}
  g_\mathrm{t,NFW}(R) = \frac{\gamma_\mathrm{t,NFW}(R)}{1 - \kappa_\mathrm{NFW}(R)} = \frac{\Sigma_{\rm crit}^{-1}\,\Delta\Sigma(R)}{1-\Sigma_{\rm crit}^{-1}\,\Sigma(R)}.
\end{equation}
Note that there exist analytic solutions for $\Sigma_\mathrm{NFW}$ and $\Delta\Sigma_\mathrm{NFW}$.

Given a set of measurements of the reduced tangential shear $\{ g_{\mathrm{t},i} \}$ at radii $\{R_i\}$, the weak-lensing mass associated with a halo is inferred by fitting the predicted shear profile to the measurements. The resulting ``weak-lensing mass'' $M_{\rm WL}$ is the mass that best reproduces the observed shear under the assumed parametric model for the halo density profile.
We define the mass bias as the ratio between this weak-lensing mass estimate and the true halo mass
\begin{equation} \label{eq:mass_bias}
  b_\mathrm{WL} \equiv \frac{M_{\rm WL}}{M_{200 \rm c}}\,.
\end{equation}

\section{Simulations} \label{sec:simulations}
The simulations used in this work are part of the \textsc{Magneticum} simulation suite\footnote{\url{http://www.magneticum.org/}} \citep{dolag16,2025arXiv250401061D} and are performed with an improved version of \texttt{P-GADGET3} \citep{beck16}, a smoothed particle hydrodynamics (SPH) code derived from \texttt{P-GADGET2} \citep{Springel:2005mi,Springel:2005nw}. These hydrodynamical simulations include a comprehensive set of physical processes, such as cooling, star formation, chemical enrichment, and feedback processes from supernovae and active galactic nuclei \citep[][and references therein]{hirschmann14, 2015ApJ...812...29T,dolag16,dolag17}.

For our analysis, we employ the set of Box1a/mr simulations. Each simulation is a medium-resolution, large-volume box with a side length of $896\,h^{-1}$Mpc. Each volume contains $1\,526^3$~dark matter particles and the same number of gas particles, corresponding to particle masses of $1.3\times 10^{10}\,h^{-1}M_\odot$ for dark matter, $2.6\times 10^{9}\,h^{-1}M_\odot$ for gas, and $6.5\times 10^{8}\,h^{-1}M_\odot$ for star particles in the WMAP7 run. 
Dark matter halos are identified using a friends-of-friends (FoF) algorithm with linking length $b_\mathrm{FoF} = 0.16$. The spherical overdensity mass of each halo is computed with the \texttt{SUBFIND} algorithm \citep{Springel:2000qu,Dolag:2008ar}, centering the halo on the particle with the minimum gravitational potential.

\begin{figure}
    \centering
    \includegraphics[width=\linewidth]{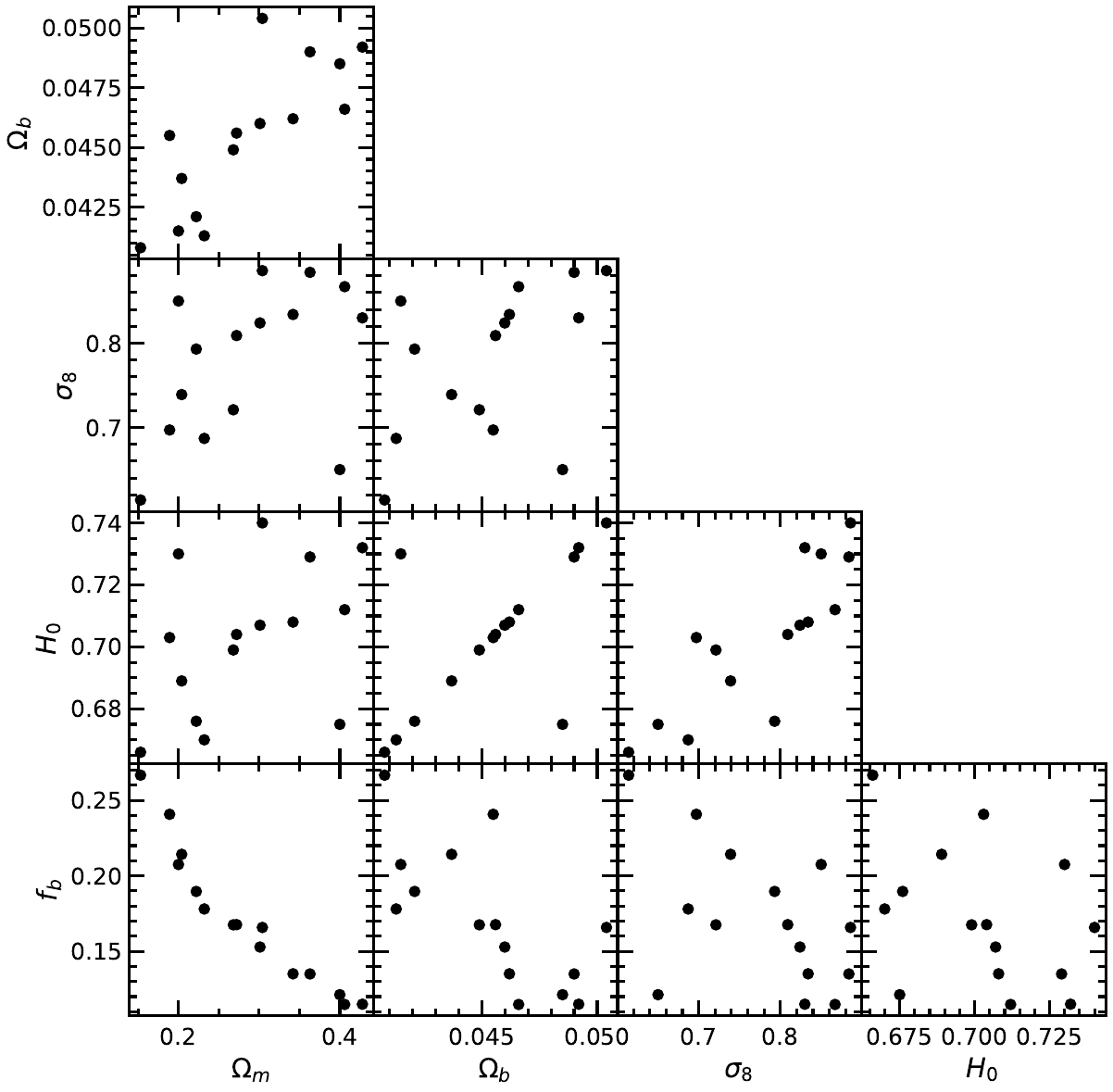}
    \caption{Combinations of cosmological parameters for the 15 Box1a runs of the \textsc{Magneticum} simulation suite.}
    \label{fig:mag_nodes}
\end{figure}

To explore cosmological dependencies, the same simulation setup is repeated for fifteen flat $\Lambda$CDM cosmologies \citep[C1--C15;][]{Singh:2019end}. The varied parameters include $\Omega_\mathrm{m}$, $\sigma_8$, $h$, and $\Omega_\mathrm{b}$, sampled via a Latin hypercube to cover the ranges $0.15 < \Omega_\mathrm{m} < 0.45$, $0.6 < \sigma_8 < 0.9$, and $0.65 < h < 0.75$, encompassing current large-scale structure constraints (see Fig.~\ref{fig:mag_nodes} and Table~\ref{tab:cosmo_params}). Thirteen of the cosmologies have a fixed $\Omega_\mathrm{b} h^{-2} \simeq 0.092$, while C3 and C13 have different values to break degeneracies between $\Omega_\mathrm{b}$ and $h$. The reference cosmology, C8, corresponds to the WMAP7 parameters for which the main set of \textsc{Magneticum} simulations was run.
To follow the time evolution, we consider simulation output at redshift $z = 0, 0.3, 0.9$.

\begin{table}
\caption{Cosmological parameters of the \textsc{Magneticum} simulations. The names are ordered by the value of $\Omega_\mathrm{m}$.}
\label{tab:cosmo_params}
\centering
\begin{tabular}{lllll}
\hline\hline
Name & $\Omega_\mathrm{m}$ & $\sigma_8$ & $f_\mathrm{b}$ & $h$\\
\hline
C1 & 0.153 & 0.614 & 0.267 & 0.666 \\
C2 & 0.189 & 0.697 & 0.241 & 0.703 \\
C3 & 0.2 & 0.85 & 0.208 & 0.73 \\
C4 & 0.204 & 0.739 & 0.214 & 0.689 \\
C5 & 0.222 & 0.793 & 0.19 & 0.676 \\
C6 & 0.232 & 0.687 & 0.178 & 0.67 \\
C7 & 0.268 & 0.721 & 0.168 & 0.699 \\
C8/WMAP7 & 0.272 & 0.809 & 0.168 & 0.704 \\
C9 & 0.301 & 0.824 & 0.153 & 0.707 \\
C10 & 0.304 & 0.886 & 0.166 & 0.74 \\
C11 & 0.342 & 0.834 & 0.135 & 0.708 \\
C12 & 0.363 & 0.884 & 0.135 & 0.729 \\
C13 & 0.4 & 0.65 & 0.121 & 0.675 \\
C14 & 0.406 & 0.867 & 0.115 & 0.712 \\
C15 & 0.428 & 0.83 & 0.115 & 0.732 \\
\hline
\end{tabular}
\end{table}

\section{Methodology} \label{sec:methods}

To quantify the weak-lensing mass bias in the simulations, we adopt a three-step procedure. First, we generate projected mass maps for the halos in each simulation box, based on which we create synthetic weak-lensing profiles in a second step. Lastly, we fit a model to the reduced tangential shear profile to infer the weak-lensing mass of each halo, which we then compare to the corresponding true halo mass in the simulation. This procedure is followed using either the gravity-only simulations or the full hydrodynamical simulations, enabling the assessment of baryonic effects on the lensing signal and thus on the lensing mass bias. Our methodology largely follows \cite{grandis21}, to which we refer the reader for additional discussion.

\subsection{Projected mass maps}

We use a parallelized pipeline to generate projected mass maps from gravity-only and hydrodynamical simulations. Halos in the gravity-only simulations are first matched to their hydrodynamical counterparts via a nearest-neighbor criterion in comoving space, taking into account periodic boundary conditions. A match is considered reliable if the separation is smaller than twice the virial radius of the halo in the gravity-only simulation. Only well-matched halos were flagged for subsequent analysis, with a mass-dependent selection ensuring a representative sampling of the halo population while keeping the computational load manageable. This procedure ensures a one-to-one match between each halo in the gravity-only simulation and its corresponding hydrodynamical halo, enabling direct comparison of their projected mass and lensing properties.

\begin{figure}
    \centering
    \includegraphics[width=\columnwidth]{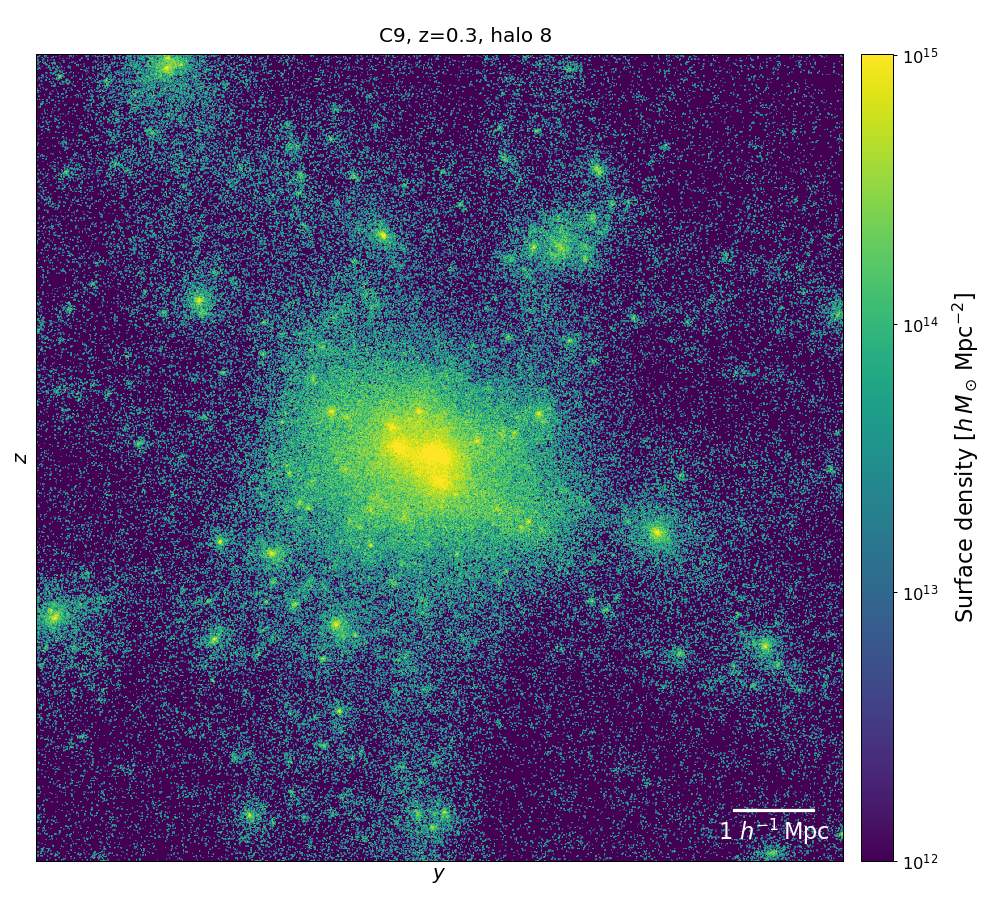}
    \caption{Projected mass map along the $x$-axis for a dark-matter halo in the C9 simulation at redshift $z=0.3$. In our analysis, we also use the projections along the $y$ and $z$-axes.}
    \label{fig:massmap}
\end{figure}

For each selected halo, projected mass maps are produced by integrating particle masses within a cylindrical volume of radius $r_{\rm max}$ and depth $l_{\rm proj}$, centered around the given cluster. We adopt $r_{\rm max}=5\,h^{-1}$Mpc and $l_{\rm proj} = 20\,h^{-1}$Mpc. Particle positions and masses -- including black hole masses in the hydrodynamical runs -- are projected onto two-dimensional grids along three orthogonal axes. A KD-tree and MPI parallelization are employed to efficiently identify particles contributing to each halo projection. An example of a projected mass map is shown in Fig.~\ref{fig:massmap}. By projecting along the three axes, each halo yields three projections. Although originating from the same halo, these projections exhibit significant variation and are treated as separate realizations, effectively tripling the size of the sample for fitting the weak-lensing mass. For the reference cosmology (i.e., WMAP7), above $M_{200\mathrm c}=1.56\times10^{14}~h^{-1}M_\odot$, we obtain 2\,532, 2\,913, and 1\,176 cluster projections at redshifts $z=0$, 0.3, and 0.9, respectively. The number of available objects depends sensitively on the underlying cosmology, leading to total numbers of cluster projections of 56\,061, 42\,120, and 17\,739 across the full set of cosmological models.

\subsection{Weak-lensing profiles}

\begin{figure}
    \centering
    \includegraphics[width=\linewidth]{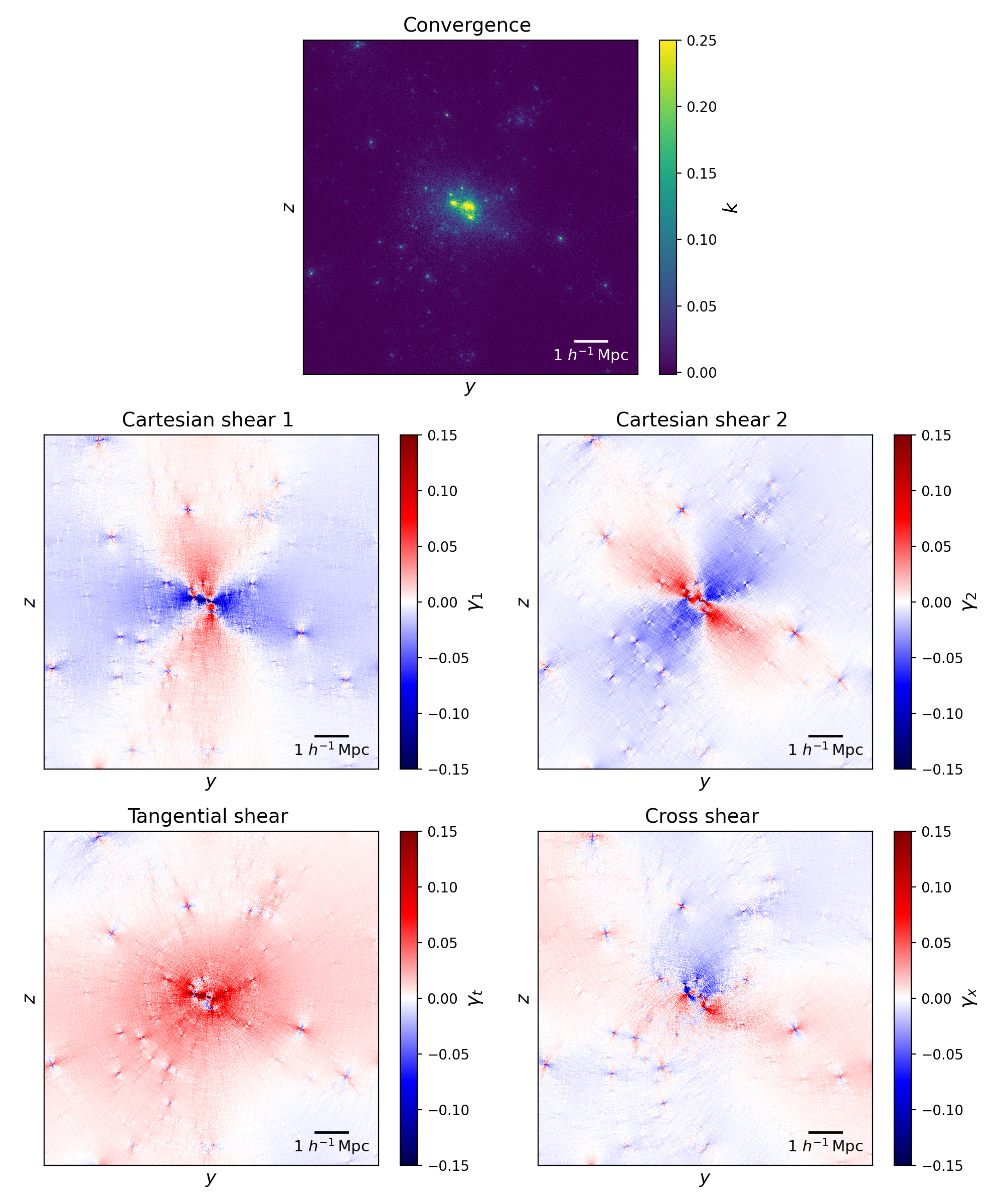}
    \caption{Maps of the convergence, the two Cartesian shear components, and the tangential and cross shear around the halo center. These maps show the same example halo as in Fig.~\ref{fig:massmap}.}
    \label{fig:2d_lensing_maps}
\end{figure}

We create weak-lensing profiles from the projected mass maps as follows. For each halo projection, we produce the surface mass density map $\Sigma$ by subtracting the contribution from the mean background density over the projection volume from the raw mass map, thus isolating the halo signal. The convergence $\kappa$ is obtained upon multiplication with $\Sigma^{-1}_\mathrm{crit}$ [see Eq.~\eqref{eq:reduced_shear}]. We apply the Kaiser--Squires method \citep{Kaiser:1992ps} to compute the shear field $\boldsymbol{\gamma}(\mathbf{x})$ via inverse Fourier transform of the two Cartesian components
\begin{equation}
\hat{\gamma}_1(\mathbf{k}) = \frac{k_1^2 - k_2^2}{k_1^2 + k_2^2}\, \hat{\kappa}(\mathbf{k})\,, 
\quad
\hat{\gamma}_2(\mathbf{k}) = \frac{2 k_1 k_2}{k_1^2 + k_2^2}\, \hat{\kappa}(\mathbf{k})\,,
\end{equation}
with the wave vector $\mathbf k$ and where $\boldsymbol{\hat\gamma}$ and $\hat\kappa$ are Fourier transforms.
The Cartesian shear components are transformed into tangential and cross shear components relative to the projected halo center
\begin{align}
&\gamma_\mathrm{t} = - \gamma_1 \cos 2\phi - \gamma_2 \sin 2\phi\,,\\
&\gamma_\times = \gamma_1 \sin 2\phi - \gamma_2 \cos 2\phi\,.
\end{align}
The tangential shear ``t'' reflects the coherent lensing distortion induced by the projected mass, whereas the cross shear ``$\times$'' is often used as a consistency check and diagnostic of residual systematics. The procedure is summarized in Fig.~\ref{fig:2d_lensing_maps}.

\begin{figure}
    \centering
    \includegraphics[width=\linewidth]{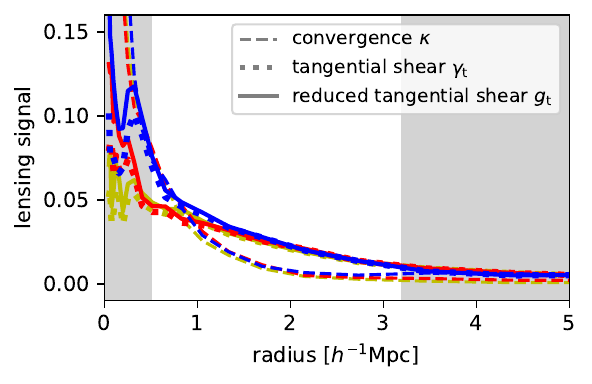}
    \caption{Radial lensing profiles for the same example halo as in Figs.~\ref{fig:massmap} and \ref{fig:2d_lensing_maps}. The three colors correspond to the profiles projected along the three Cartesian axes.}
    \label{fig:1d_lensing_profile}
\end{figure}

Radial and tangential binning is then applied to the convergence and the tangential shear fields, from which the reduced tangential shear field is computed. The radial tangential shear profile is then obtained by azimuthal averaging,
\begin{equation}
  \langle g_\mathrm{t}\rangle(R) = \int \mathrm d\phi\frac{\gamma_\mathrm{t}(R,\phi)}{1-\kappa(R,\phi)}.
\end{equation}
Note that one first computes the two-dimensional reduced shear field and then performs the azimuthal averaging (because the integrand is nonlinear, these two operations do not commute).
The final output for each projection consists of the convergence, tangential shear, and reduced tangential shear profiles as functions of projected radius (see Fig.~\ref{fig:1d_lensing_profile} for an example).

Note that we did not include or account for miscentering in this analysis, as accounting for offsets between the true halo center and the assumed center introduces an additional layer of complexity that is beyond the scope of the present work. Calibrations of the lensing mass bias used in cosmological analyses do of course account for the effect of miscentering \citep{grandis21, bocquet24a, grandis24}.

\subsection{Weak-lensing mass fit}

In the final step of the analysis, we fit the parametric lensing model defined in Sect.~\ref{sec:theory} to the measured reduced tangential shear profiles to infer a weak-lensing mass for each halo projection. We recompute the critical surface mass density $\Sigma_\mathrm{crit}$ for each snapshot and for each cosmology, so that the cosmology-dependent change of angular diameter distances is accounted for. Then, following Eq.~\eqref{eq:reduced_shear} and assuming the NFW mass profile and a selection of models for the halo concentration (see Sec.~\ref{sec:results}), we compute the model profile $g_\mathrm{t}^\mathrm{model}(R, M_\mathrm{WL})$ as a function of lensing mass $M_\mathrm{WL}$.

We restrict our analysis to the radial range
\begin{equation}
  0.5<R\,[h^{-1}\mathrm{Mpc}]<3.2\,(1+z)^{-1}
\end{equation}
to exclude the cluster central region and to avoid larger separations beyond the 1-halo term regime, as motivated by \cite{grandis21} and applied in analyses of DES lensing around SPT, eROSITA, and \textit{Planck} clusters \citep{bocquet24a, grandis24, aymerich25}.
We define the weighted squared difference between the reduced shear $g_\mathrm{t}^\mathrm{sim}(R)$ measured in the simulation and the model prediction
\begin{equation}
  \chi^2(M_\mathrm{WL}) = \sum_i \left[ g_\mathrm t^\mathrm{model}(R_i, M_\mathrm{WL}) - g_\mathrm t^\mathrm{sim}(R_i) \right]^2 w_i^2,
\end{equation}
with weights $w$. We apply uniform weighting in projected area and the weights thus correspond to the areas of the angular bins.\footnote{Because we are only interested in the best-fitting mass, we do not explicitly assume or model any measurement errors.} We obtain the lensing mass $M_\mathrm{WL}$ for each halo by minimizing $\chi^2$ using a standard minimization algorithm (amoeba). While estimators based on the full posterior distribution may be less prone to outliers and may reduce the scatter in the $M_\mathrm{WL}-M_\mathrm{halo}$ relation, our idealized, noiseless setup justifies the use of the minimum-$\chi^2$ estimator \citep[see, e.g.,][]{Euclid:2025Ingoglia}.

\section{Results} \label{sec:results}

\begin{figure}
  \centering
  \includegraphics[width=\linewidth]{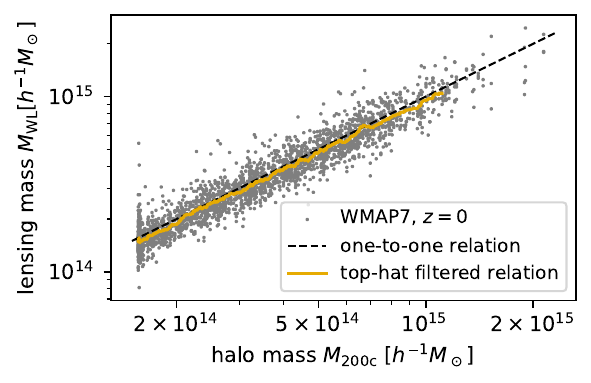}
  \caption{Relationship between the halo mass and the lensing mass computed from the reduced shear profile for each halo. The top-hat filtered relation qualitatively follows the one-to-one relation, implying that the simplified model for the cluster shear is appropriate.}
  \label{fig:MwlMhalo}
\end{figure}

\begin{figure}
  \centering
  \includegraphics[width=\linewidth]{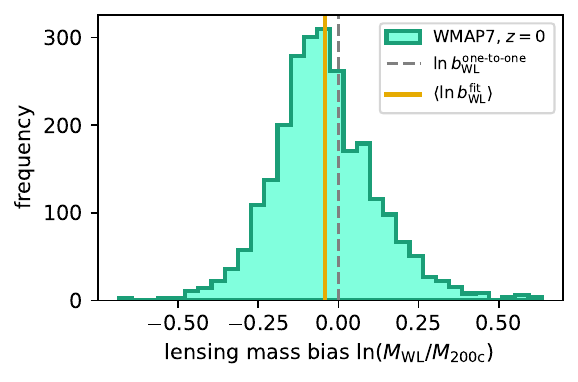}
  \caption{The distribution of ratios between the lensing mass and the halo mass is well described by a lognormal distribution. The uncertainty on the mean is too small to be displayed in this figure.}
  \label{fig:MwlMhalo_hist}
\end{figure}

We follow the procedures described in the previous sections and obtain a weak-lensing mass $M_\mathrm{WL}$ for each halo projection in our synthetic dataset. Figure~\ref{fig:MwlMhalo} shows the $M_\mathrm{WL}-M_\mathrm{halo}$ relation for the fiducial cosmology at $z=0$. The visual impression suggests that the cluster weak-lensing mass bias $b_\mathrm{WL}\equiv M_\mathrm{WL}/M_\mathrm{halo}$ is close to unity.
In Fig.~\ref{fig:MwlMhalo_hist}, we show the histogram of the logarithms of the lensing mass bias for the same cosmology and redshift. There is 16\% scatter in the relation, and the mean is offset from the one-to-one relation by a few percent: We recover $\langle\ln b_\mathrm{WL}^{c=3.5,\text{ C8}}\rangle=-0.042\pm0.004$.

In what follows, we investigate the dependence of the mass bias on the underlying cosmology. We find that there is more than one effect at play. Therefore, we decompose the net variations into the individual contributions.

\subsection{Cosmology dependence of the weak-lensing mass bias in gravity-only simulations}
\label{subsec:gravonly}

\begin{figure*}
  \centering
  \begin{minipage}[t]{.48\textwidth}\vspace{0pt}
    \centering
    \includegraphics[width=\linewidth]{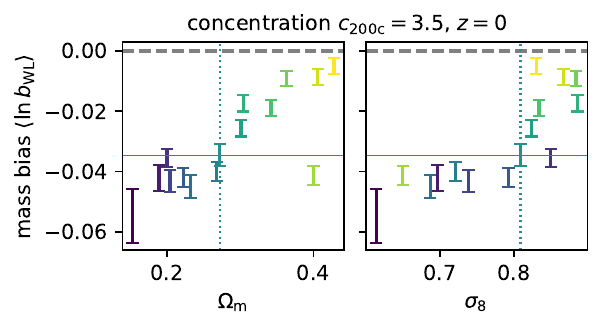}
    \caption{Cluster lensing mass bias in gravity-only simulations, assuming a fixed concentration $c_{200\mathrm{c}}=3.5$. Error bars show the mean bias $\pm1\sigma$ and are color-coded according to their $\Omega_\mathrm{m}$ value. The horizontal line indicates the mass bias in our fiducial WMAP7 cosmology, the vertical dotted lines indicate the WMAP7 cosmological parameters. The bias increases with increasing values of $\Omega_\mathrm{m}$ and $\sigma_8$. Across the range of cosmologies we probe, the intrinsic standard deviation in the logarithmic mass bias is 0.014.}
    \label{fig:gravonly_fixed_c}
  \end{minipage}\hfill
  \begin{minipage}[t]{.48\textwidth}\vspace{0pt}
    \centering
    \includegraphics[width=\linewidth]{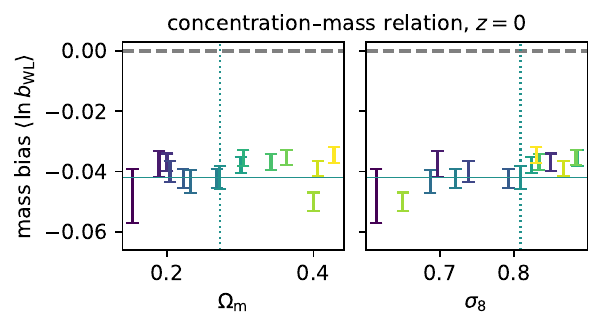}
    \caption{Cluster lensing mass bias in gravity-only simulations, assuming a cosmology-dependent concentration--mass relation \citep{Diemer:2019vmz}. Error bars show the mean bias $\pm1\sigma$ and are color-coded according to their $\Omega_\mathrm{m}$ value. The horizontal line indicates the mass bias in our fiducial WMAP7 cosmology, the vertical dotted lines indicate the WMAP7 cosmological parameters. In comparison with Fig.~\ref{fig:gravonly_fixed_c}, the more realistic concentration absorbs the cosmology dependence of the halo profiles. The intrinsic standard deviation in the logarithmic mass bias is reduced to 0.0026 and we observe no clear trend with $\Omega_\mathrm{m}$ or $\sigma_8$.}
    \label{fig:gravonly_c-M}
  \end{minipage}
\end{figure*}

First, we consider the simplest model and set the concentration to a fixed value $c_{200\mathrm{c}}=3.5$ that is representative for massive halos \citep[e.g.,][]{child18}. We show the recovered mass bias as a function of $\Omega_\mathrm{m}$ and $\sigma_8$ in Fig.~\ref{fig:gravonly_fixed_c}. The inverse-variance weighted mean bias across all cosmologies
\begin{equation}
  \langle \ln b_\mathrm{WL}^{c=3.5,\text{ gravity-only}}\rangle_\text{all cosmologies} = -0.0258\pm0.0008
\end{equation}
differs from unity because our halo mass model is not perfect. This fact has been shown previously \citep[e.g.,][]{Becker:2010xj} and cluster lensing analyses account for this bias. The lensing mass bias for our reference WMAP7 cosmology, $\ln b_\mathrm{WL}^\mathrm{WMAP7} = -0.035\pm0.004$ is very close to the mean bias across all cosmologies. However, from the figure it is clear that assuming a single value for the mass bias would mean ignoring the trend with $\Omega_\mathrm{m}$ and $\sigma_8$. Indeed, the individual values for the logarithmic bias vary from $-0.005\pm0.003$ (C15; high $\Omega_\mathrm{m}$ and high $\sigma_8$) to $-0.055\pm0.009$ (C1; low $\Omega_\mathrm{m}$ and low $\sigma_8$). We define the intrinsic standard deviation as
\begin{equation}
  s_\mathrm{int} \equiv \sqrt{\frac{\sum_{i=\mathrm{C1}}^\mathrm{C15} \Bigl[\bigl(\langle\ln b_{\mathrm{WL}}\rangle_i - \langle \ln b_\mathrm{WL}\rangle_\text{all cosmo.}\bigr)^2/\sigma_i^2\Bigr]}{\sum_{i=\mathrm{C1}}^\mathrm{C15}\sigma_i^{-2}} - \frac{N}{\sum_{i=\mathrm{C1}}^\mathrm{C15} \sigma_i^{-2}}},
\end{equation}
with $N=15$ and obtain
\begin{equation}
  s_\mathrm{int}^{c=3.5,\text{ gravity-only}}=0.014.
\end{equation}
This spread in the recovered values of the bias can be interpreted as an additional 1--2\% uncertainty in mass when taking the WMAP7 value as the reference. We note that this uncertainty is small but not irrelevant compared to the overall systematic and statistical uncertainties in current cluster weak-lensing mass calibration analyses \citep[e.g.,][]{bocquet24a, grandis24}.

Next, we improve our halo mass model by adopting the concentration model by \cite{Diemer:2019vmz} that depends on the halo peak height and on the effective slope of the matter power spectrum to capture the dependence on halo mass, redshift, and cosmology. In this case, as can be seen in Fig.~\ref{fig:gravonly_c-M}, the scatter across cosmologies is reduced and the figure does not suggest any obvious residual trend of the bias with $\Omega_\mathrm{m}$ or $\sigma_8$. We report an intrinsic standard deviation
\begin{equation}
  \label{eq:sintDJ19gravonly}
  s_\mathrm{int}^\text{Diemer\&Joyce19, gravity-only}=0.0027
\end{equation}
that is over five times smaller than for the model with fixed concentration. Our conclusion is that, to achieve sub-percent robustness, the cluster lensing mass model should contain a cosmology-dependent halo concentration.

We also consider other concentration--mass relations. \cite{duffy08} account for the mass dependence of concentration, but as their relation does not account for any dependence with cosmology, we observe a behavior that is similar to the fixed-concentration scenario shown in Fig.~\ref{fig:gravonly_fixed_c}. We recover
\begin{equation}
  s_\mathrm{int}^\text{Duffy+08}=0.015.
\end{equation}
\cite{child18} propose a parametrization of concentration that depends on the nonlinear mass scale $M_\star$ and thus on the underlying cosmology. Indeed, using their relation improves the variation of the logarithmic mass bias by about a factor of two compared to the results obtained with a fixed concentration or the \cite{duffy08} relation, and we obtain
\begin{equation}
  s_\mathrm{int}^\text{Child+18}=0.005.
\end{equation}
However, especially for low values of $\Omega_\mathrm{m}$, the results obtained with the \cite{Diemer:2019vmz} model appear to be the least dependent on cosmology.

\subsection{Cosmology dependence of the weak-lensing mass bias in hydrodynamical simulations}
\label{subsec:hydro}

We now consider the full-physics hydrodynamical \textsc{Magneticum} runs. For plausible strengths of feedback, cluster-scale halos are not disrupted. However, due to the complex interplay between cooling and feedback, the halo mass profiles are altered compared to the gravity-only (and dark-matter only) scenario. Due to the change in profile, the overdensity radius $r_{200\mathrm{c}}$ is changed and with it, the enclosed mass $M_{200\mathrm{c}}$. Indeed, the halo masses in the \textsc{Magneticum} and Illustris-TNG hydrodynamical runs differ from the masses in the gravity-only counterpart simulations by a few percent \citep[e.g.,][]{castro21, grandis21}.

\begin{figure*}
  \centering
  \includegraphics[width=\linewidth]{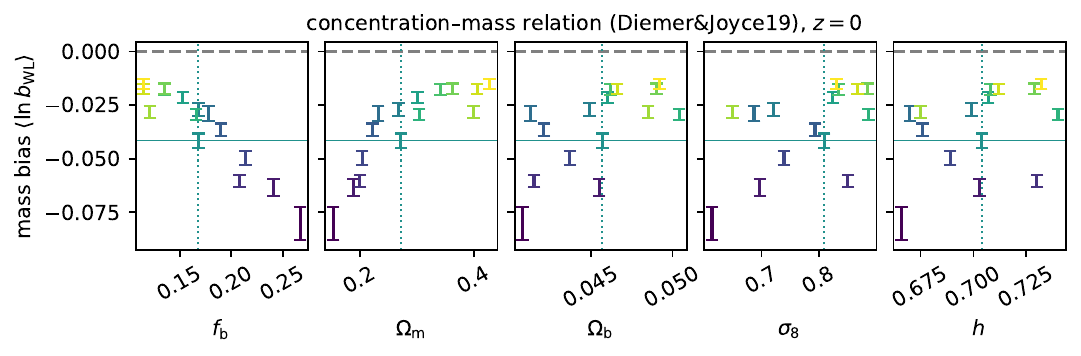}
  \caption{Cluster lensing mass bias in hydrodynamical simulations, assuming a concentration--mass relation that varies as a function of cosmology \citep{Diemer:2019vmz}. Error bars show the mean bias $\pm1\sigma$ and are color-coded according to their $\Omega_\mathrm{m}$ value. The horizontal line indicates the mass bias in our fiducial WMAP7 cosmology, the vertical dotted line indicates the WMAP7 cosmological parameters.}
  \label{fig:hydro_c-M}
\end{figure*}

Motivated by our findings from the previous section, we adopt the \cite{Diemer:2019vmz} concentration model and fit for the relationship between the lensing mass and the halo mass as measured in the hydrodynamical simulation.
In Fig.~\ref{fig:hydro_c-M}, we show the dependence of the mass bias in hydrodynamical simulations with the cosmological parameters. We report an intrinsic standard deviation in $\ln b_\mathrm{WL}$
\begin{equation}
  \label{eq:s^hydro}
  s_\mathrm{int}^\text{Diemer\&Joyce19, hydro} = 0.015,
\end{equation}
with values ranging from $-0.080\pm0.008$ (C1) to $-0.015\pm0.002$ (C15). The inverse-variance weighted mean bias is 
\begin{equation}
  \label{eq:mean^hydro}
  \langle\ln b_\mathrm{WL}^\text{Diemer\&Joyce19, hydro}\rangle_\text{all cosmologies} =-0.0298\pm0.0007.
\end{equation}
From Fig.~\ref{fig:hydro_c-M}, it appears as though there is a strong dependence of the mass bias with $\Omega_\mathrm{m}$. Conversely, the parameters $\Omega_\mathrm{b}$, $\sigma_8$, and $h$ do not seem to have a strong impact. However, the figure also suggests that the trend of the bias with $\Omega_\mathrm{m}$ is almost exactly inverse to the trend with the baryon fraction $f_\mathrm{b}$. Because our results from Sect.~\ref{subsec:gravonly} did not suggest such a strong trend with $\Omega_\mathrm{m}$, and because all 15 parameter combinations $\Omega_\mathrm{m}-f_\mathrm{b}$ of the \textsc{Magneticum} simulations are located along a narrow band in parameter space (see Fig.~\ref{fig:mag_nodes}), we conclude that the baryon fraction is responsible for driving the evolution of the mass bias. This is not unexpected, because the global baryon fraction also modulates the local baryon fraction in massive halos and thus the strength of baryonic effects on the halo mass profiles. In conclusion, we are seeing the impact of varying strengths of baryonic effects rather than the impact of changing background cosmologies.

\begin{figure}
  \includegraphics[width=\linewidth]{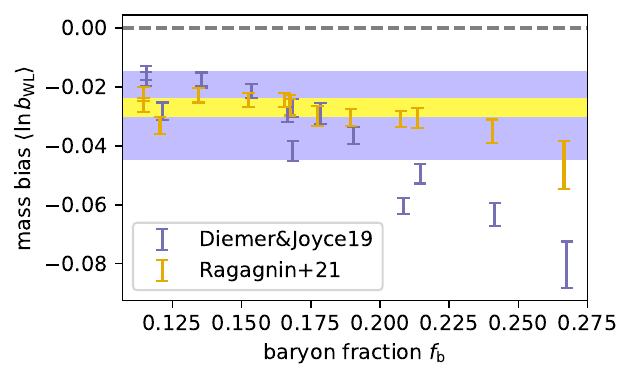}
  \caption{Cluster lensing mass bias obtained using two cosmology-dependent halo concentration models. The shaded regions show the respective mean $\pm$ the intrinsic standard deviation among the data points. The \cite{ragagnin21con} model explicitly depends on $\Omega_\mathrm{b}$ and $\Omega_\mathrm{m}$ and absorbs essentially all of the evolution of the bias with the baryon fraction $f_\mathrm{b}\equiv \Omega_\mathrm{b}/\Omega_\mathrm{m}$.}
  \label{fig:Ragagnin}
\end{figure}

Our finding implies that an improved halo mass model should explicitly depend on the baryon fraction. Therefore, we now consider a concentration--mass relation that is calibrated on the same suite of \textsc{Magneticum} simulations we also use \citep{ragagnin21con}. In this relation, concentration is modeled as a power law in halo mass and scale factor $(1+z)^{-1}$, and the amplitude, mass dependence, and redshift dependence are each modeled as power laws in $\Omega_\mathrm{m}$, $\Omega_\mathrm{b}$, $H_0$, and $\sigma_8$. Using this model, we fit for the lensing mass bias and obtain an intrinsic standard deviation
\begin{equation}
  s_\mathrm{int}^\text{Ragagnin+21} = 0.0030
\end{equation}
that is about five times smaller than for the \cite{Diemer:2019vmz} model. In Fig.~\ref{fig:Ragagnin}, we show the lensing mass bias obtained for the two different concentration--mass relations. We note that because \cite{ragagnin21con} calibrated their model using the same hydrodynamical simulations as we do, there is no guarantee that the residuals would be similarly small when using other hydrodynamical simulations with different prescriptions for baryonic feedback. However, such a test is beyond the scope of this work.

\subsection{Matching weak lensing in hydrodynamical simulations to the gravity-only halo mass function}

Feedback effects in halos tend to push some of the material from the inner part to larger radii. As a result, the halo mass profile changes. For our purposes, this has two consequences: i) the mass $M_{200\mathrm{c}}$ decreases by a few percent, and ii) the lensing profile is altered. The latter effect is causing a shift in the inferred lensing mass $M_\mathrm{WL}$. However, at least for massive halos in the cluster regime, the underlying halo is not disrupted. Baryonic feedback effects can thus be modeled as small perturbations around a gravity-only model. For the purpose of cluster cosmology, \cite{grandis21} proposed to calibrate the lensing mass bias between the halo mass as measured in a gravity-only simulation and the realistic lensing profile as measured in a hydrodynamical simulation. This is achieved by matching halos between gravity-only and full-physics hydrodynamical simulations that are run with identical initial conditions. The advantage of this approach is that the halo mass function remains the function calibrated in gravity-only simulations, while the impact of hydrodynamical effects and their uncertainties are folded into the $M_\mathrm{WL}$--$M_\mathrm{halo}$ relation. Recent state-of-the-art cluster abundance cosmology analyses of eROSITA, \textit{Planck}, and SPT clusters have adopted that approach \citep{SPT:2024qbr, Ghirardini:2024yni, aymerich25}.

We now investigate the cosmological impact on the relation between lensing mass in full-physics hydrodynamical simulations and the halo mass in their gravity-only counterpart. The results we obtain are very similar to the ones we present in the previous subsection~\ref{subsec:hydro} and in Fig.~\ref{fig:hydro_c-M}. We report an inverse-variance weighted mean bias across all cosmologies
\begin{equation}
  \langle\ln \left(\frac{M_\mathrm{WL}^\text{hydro, Diemer\&Joyce19}}{M_\mathrm{halo}^\mathrm{gravity-only}}\right)\rangle_\text{all cosmologies}=-0.0297\pm0.0007
\end{equation}
and an intrinsic standard deviation in $\langle\ln b_\mathrm{WL}\rangle$
\begin{equation}
  s_\mathrm{int}\left( \frac{M_\mathrm{WL}^\text{hydro, Diemer\&Joyce19}}{M_\mathrm{halo}^\mathrm{gravity-only}}\right) = 0.015,
\end{equation}
closely matching the results in Eqs.~\eqref{eq:s^hydro} and \eqref{eq:mean^hydro}. We thus conclude that the observations made in the previous subsection~\ref{subsec:hydro}, where we evaluated the bias between the lensing mass and halo mass from hydrodynamical simulations, also applies to the approach we consider here, where we compare the lensing mass from hydrodynamical simulations with the halo mass from gravity-only simulations.

\subsection{Evolution with cluster redshift}

\begin{figure}
  \centering
  \includegraphics[width=\linewidth]{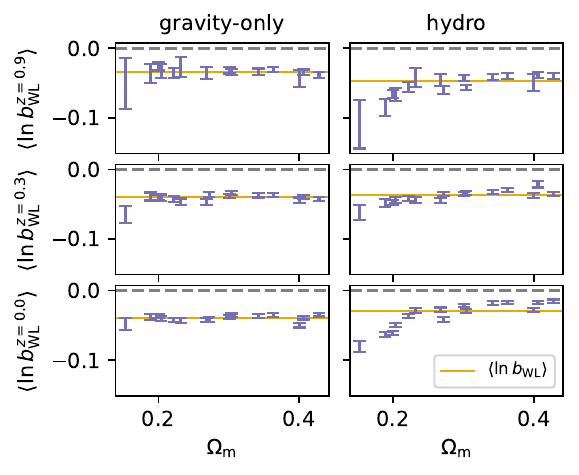}
  \caption{The cluster weak-lensing mass bias as a function of $\Omega_\mathrm{m}$ at different redshifts. We consider the concentration--mass relation by \cite{Diemer:2019vmz}. The two bottom panels correspond to the $\Omega_\mathrm{m}$-panels in Figs.~\ref{fig:gravonly_c-M} and \ref{fig:hydro_c-M}. In this figure, the orange line shows the mean logarithmic mass bias across all cosmologies at the given redshift. The details of the redshift-evolutions appear non-trivial, but the qualitative picture does not change with redshift.}
  \label{fig:redshift}
\end{figure}

Thus far, we have restricted our analysis and discussion to redshift $z=0$. We now repeat the gravity-only and hydrodynamical analyses from subsections~\ref{subsec:gravonly} and \ref{subsec:hydro} and consider redshifts $z\in\{0,0.3,0.9\}$. We assume the \cite{Diemer:2019vmz} concentration model. Our results are shown in Fig.~\ref{fig:redshift}.

In the gravity-only case, the mean logarithmic mass bias $\langle\ln b_\mathrm{WL}\rangle$ evolves from $-0.034$ at $z=0.9$ to $-0.039$ at redshift zero. The intrinsic standard deviation evolves from 0 at $z=0.9$ to 0.0026 [as in Eq.~\eqref{eq:sintDJ19gravonly}].\footnote{At $z=0.9$, the spread across all data points is less than the variation expected from the individual error bars. We thus measure a negative intrinsic variance $s_\mathrm{int}^2<0$.} Our conclusions from subsection~\ref{subsec:gravonly} thus remain unchanged: Assuming the \cite{Diemer:2019vmz} concentration--mass relation, we observe no strong dependence of the cluster lensing mass bias with cosmology.

We now consider the hydrodynamical simulations as in subsection~\ref{subsec:hydro}. As shown in Fig.~\ref{fig:redshift}, the qualitative trend of the bias with $\Omega_\mathrm{m}$ is similar at all redshifts. The mean logarithmic mass bias $\langle\ln b_\mathrm{WL}\rangle$ evolves from $-0.048$ at redshift $z=0.9$ to $-0.0298$ at $z=0$ [as reported in Eq.~\eqref{eq:mean^hydro}]. The intrinsic standard deviation at higher redshifts is smaller than at $z=0$ [see Eq.~\eqref{eq:s^hydro}], and we obtain $0.009$ at $z=0.9$ and $0.006$ at $z=0.3$. We conclude that our discussion in subsection~\ref{subsec:hydro} applies also to higher redshifts: a better model for cluster lensing will have to explicitly account for baryonic effects.

\section{Conclusions} \label{sec:conclusions}

In this work, we investigate the impact of cosmology on cluster lensing. The cosmology-dependence of the lens--source geometry can be easily accounted for by including the cosmologically correct angular diameter distances that set the critical surface mass density, see Eq.~\eqref{eq:Sigma_crit}. Here, we consider the impact that cosmology has on the halos themselves. We use a suite of 15 \textsc{Magneticum} cosmological hydrodynamical simulations and extract synthetic lensing (shear) measurements (see Figs.~\ref{fig:massmap}--\ref{fig:1d_lensing_profile}). In total, we produce and analyze 115\,920~synthetic cluster shear maps. We measure the lensing mass $M_\mathrm{WL}$ in each halo map assuming an NFW mass profile and consider the radial range $0.5<R/(h^{-1}\mathrm{Mpc})<3.2\,(1+z)^{-1}$ (see Fig.~\ref{fig:MwlMhalo}). Finally, we investigate the dependence of the weak-lensing mass bias $b_\mathrm{WL}\equiv M_\mathrm{WL}/M_\mathrm{halo}$ with cosmology. Our results are summarized as follows:
\begin{itemize}
  \item In a gravity-only universe, and when assuming a concentration--mass relation that does not vary with cosmology (e.g., $c=3.5$), there is a clear trend of the lensing mass bias with cosmology, see Fig.~\ref{fig:gravonly_fixed_c}. The intrinsic standard deviation in the logarithmic mass bias across cosmologies is $s_\mathrm{int}=0.014$.
  \item Still in the gravity-only scenario, using the cosmology-dependent concentration--mass relation by \cite{Diemer:2019vmz} significantly reduces the cosmological dependence of the mass bias (see Fig.~\ref{fig:gravonly_c-M}). The remaining intrinsic standard deviation in the mass bias across cosmologies is small ($s_\mathrm{int}=0.0027$).
  \item In the hydrodynamical simulations, there is a clear trend of the mass bias with the universal baryon fraction (see Fig.~\ref{fig:hydro_c-M}). The concentration--mass relation that was calibrated using the same \textsc{Magneticum} simulations we also use here \citep{ragagnin21con} absorbs most of this trend (see Fig.~\ref{fig:Ragagnin}).
\end{itemize}

Some of the recent cosmological analyses rely on a cluster lensing model with a fixed concentration \citep{bocquet24a, SPT:2024qbr,  aymerich25}. As we show in this work, this choice is not optimal, but the resulting changes in the lensing mass bias $\Delta\ln b_\mathrm{WL}$ are small compared to the overall systematic and statistical uncertainties in cluster lensing. Yet, our results suggest that using a cosmology-dependent model for the halo concentration is more appropriate. The eROSITA analyses relies on the \cite{ragagnin21con} relation and should therefore be immune to cosmology-dependent and baryonic physics dependent changes of the lensing mass bias over the range of cosmologies and baryonic physics models covered by the \textsc{Magneticum} mr simulations \citep{grandis24, Ghirardini:2024yni}.

We report a dependence of the lensing mass bias on the universal baryon fraction. However, the impact of baryonic effects in halos is set by the interplay between the availability of baryons and the strength of feedback. To discriminate between these two effects, one would require a suite of simulations where both the cosmological parameters (including $f_\mathrm{b}$) and the feedback prescriptions are varied. The \textsc{Magneticum} simulations do not offer this possibility, but suites like FLAMINGO might prove useful in this context \citep{Schaye_flamingo_23}. Finally, while we consider the universal baryon fraction $f_\mathrm{b}$ in this work, we expect the local baryon fraction within some radius [such as $f_\mathrm{b}(r_{500\mathrm{c}})$] to be more closely linked to the baryonic processes inside halos \citep[e.g.,][]{vandaalen20}. We recommend that future analyses adopt more sophisticated models for the halo concentration that depend on cosmology and on the strength of baryonic feedback.

For simplicity, the analysis we perform and present here ignores several complications that arise in the analysis of real data (shape noise, shear and photo-$z$ calibration, cluster member contamination of the shear signal, miscentering). Our goal is to isolate the impact of cosmology and we make the implicit assumption that none of these additional effects couple to the underlying cosmology. Due to these simplifications, the bias values we obtain are not meant to be used to obtain lensing-based cluster mass estimates.  We hope that the observations and recommendations we make here will help inform the cluster lensing analysis frameworks of upcoming studies.

\begin{acknowledgements}
We acknowledge support by LMU Munich and the Excellence Cluster ORIGINS, funded by the Deutsche Forschungsgemeinschaft (DFG, German Research Foundation) under Germany's Excellence Strategy -- EXC 2094 -- 390783311. 
KD acknowledges support by the COMPLEX project from the European Research Council (ERC) under the European Union’s Horizon 2020 research and innovation program grant agreement ERC-2019-AdG 882679.
The calculations for the hydrodynamical simulations were carried out at the  Leibniz Supercomputer Center (LRZ) under the project pr83li (Magneticum). 
\end{acknowledgements}

\bibliographystyle{aa} 
\bibliography{biblio}

\end{document}